\documentclass{article}
\usepackage{amsmath}
\usepackage{graphicx}
\usepackage{float}

\input oejv.sty

\begin{document}
\newcommand{\mdot}{\overset{^{\rm m}}{.}}
\newcommand{\pdot}{\overset{^{\rm d}}{.}}

\OEJVhead{July 2013}

\begin{center}
\textbf{Nine New Variable Stars in Camelopardalis}

Natalia A. Virnina

Department of High and Applied Mathematics, Odessa National Maritime University, Ukraine, virnina@gmail.com
\end{center}

\textbf{Abstract:} Nine new short period variable stars have been discovered in the direction of the open cluster Cr464 in Camelopardalis. The field was observed using Tzec Maun Observatory's telescope AP-180. Two new variable stars were classified as pulsating stars (RRab and RRc types). The other seven stars are binary systems. One of them is of ELL-type, four binaries were recognized as EW-type, and two systems are rather short period EA-type binaries.
All new variables were registered in the VSX catalogue.

\bigskip
\textbf{1. Observations}
\bigskip

The field centered on $R.A.=05^h 30^m 19.1^s, Dec=+72^\circ 57' 30''$ has been chosen for searching for new variable stars due to a lack of known variables in it. Unfiltered CCD observations were made on 12 nights between 22.02.2010 and 19.03.2010 using remotely controlled AP-180 Astro-Physical refractor ($D=180{\rm mm}, F=1317{\rm mm}$) of Tzec Maun Observatory (New Mexico, USA), and CCD camera SBIG STL-11000. The field of view was $87.5' \times 58.3'$, the resolution $2.627''$/pixel. We had no opportunity to use any photometric filter. However, the maximum quantum efficiency of the camera sensor is close to the standard $R_c$-band (Virnina 2010). Exposure duration was 180 sec on all nights. All images were processed with dark frames and flat fields. Altogether we collected 438 images, fit for photometric study.

\bigskip
\textbf{2. Comparison stars}
\bigskip

During three nights the field of view overlapped the field of RR Cam. Thus, five reference stars (listed in the Table 1) were chosen to create a "secondary standard". The magnitudes of these five stars were measured by APASS (AAVSO). Three other stars close to the center of our field were measured photometrically relative to the ensemble of five reference stars on 46 best images. $R_c$-band magnitudes were used for photometry processing. These three stars were used as comparison stars for ensemble measuring of all new variables. The information about comparison stars is listed in the Table 2. Location of these stars is shown in the Fig. 1.

\bigskip
\textbf{3. Searching for new variable stars}
\bigskip

We searched for new variable stars using the software package C-Munipack (Motl 2007), based on the SExtractor routine (Bertin \& Arnouts 1996). To identify possible variable stars among all stars in the field with $SNR>3$, we created a $RMS$-scatter versus mean magnitude diagram. The fainter objects tend to have larger scatter. Those objects, which deviate from this relation, are good candidates to be variable stars. False positives could also be caused by either the presence of a close companion, or a bright star in the vicinity of the object, or a galaxy, or image defect etc.
Nine new short-period ($P<1d$) variable stars were discovered in the searched field. Two of them were recognized as pulsating variables, the other seven - as close binary systems. We've also searched for long period variables, using the images from different sets, but haven't found any.
All coordinates of new variable stars were taken from the USNO-B1.0 catalog (Monet et al. 2003). Once the photometric parameters were determined, new variable stars were submitted and approved in the VSX catalogue of variable stars, maintained by AAVSO.

\bigskip
\textbf{4. Photometric parameters of new variable stars}
\bigskip

The time-dependent light curves (HJD based on UTC) of newly discovered variable stars were searched for periodic variation of brightness using the method by Lafler \& Kinman (1965) implemented in the Peranso v2.20 software (Vanmunster 2007). When approximate periods were determined, the FDCN program (Andronov 1994, 2003) was used, which computes the coefficients of the statistically optimal trigonometric polynomials using the least squares method routine and differential corrections for the period. It determines the period value and initial epoch with corresponding errors estimates.
The coordinates, USNO-B1.0 and VSX numbers of new variable stars are listed in Table 3, while all photometric parameters with their errors, and presumed types of variability are summarized in the Table 4. The precision of the photometric measurements could be evaluated as a standard deviation between comparison stars: $\sigma(C_1-C_2)=0\mdot023$, $\sigma(C_2-C_3)=0\mdot029$, $\sigma(C_3-C_1)=0\mdot031$. 

\bigskip

\bigskip
\textit{4.1 VSX J052543.6+722840}

The period of USNO-B1.0 1624-0065083=VSX J052543.6+722840 is $P=0\pdot64221\pm0\pdot00013$. To determine the depths of minima, we calculated the coefficients of approximating algebraic polynomials using MCV software (Andronov \& Baklanov 2004). The optimal degrees of the polynomials were $s=4$ for both minima. This approximation yielded that minima are nearly equal (min$_{\rm I}=16\mdot035\pm0\mdot026$, min$_{\rm II}=15\mdot987\pm0\mdot021$). The location of this variable is marked in Fig. 2, and the phase curve is shown in Fig. 3. Despite the fact that the period is rather short, but taking into account that the eclipses are narrow (their corresponding durations are 0.135$\phi$ and 0.179$\phi$), we classified this binary as EA-type (Algol). 

\bigskip
\textit{4.2 VSX J052825.5+732114}

USNO-B1.0 1633-0056300 = VSX J052825.5+732114 is another binary star. The location of this star is shown in Fig. 4. Its period is $P=0\pdot41535\pm0\pdot00002$. The depths of minima were determined using MCV software as was described above. The magnitudes in minima are nearly equal (min$_{\rm I}=14\mdot386\pm0\mdot003$, min$_{\rm II}=14\mdot383\pm0\mdot005$), or in fact equal within the errors, and one may assume that the primary minimum is actually the secondary one and vice versa. Taking into account the shape of the phase curve (Fig. 5) and other photometric characteristics of this binary system, we classified it as EW-type binary.

\bigskip
\textit{4.3 VSX J052916.7+723300}

The position of USNO-B1.0 1625-0065763 = VSX J052916.7+723300 is indicated in Fig. 6. The phase curve of this variable star is not covered completely (Fig. 7). However, we found the period of variability $P=0\pdot91252\pm0\pdot00015$. The amplitude of this star is rather small: max $=12\mdot177\pm0\mdot002$, min $=12\mdot292\pm0\mdot004$, $\Delta_m=0\mdot115\pm0\mdot004$. As there are two minima and two maxima on the phase curve, we assumed that this variable star is a binary system. However, there is no evidence that this binary is eclipsing, the curve is sinusoidal. In such systems the variability is caused only by the geometry of the components. Thus, we classified this variable as ELL-type star.

\bigskip
\textit{4.4 VSX J053044.4+725113}

Another EA system, USNO-B1.0 1628-0064829 = VSX J053044.4+725113, has an even shorter period $P=0\pdot33648\pm0\pdot00002$. The location of this star is shown in Fig. 8. To determine magnitudes in maximum and minima, we approximated corresponding parts of the phase curves by the algebraic polynomials of the optimal degree $s=4$ using MCV software: max $=14\mdot901\pm0\mdot006$, min$_{\rm I}=15\mdot476\pm0\mdot010$, min$_{\rm II}=15\mdot369\pm0\mdot010$. On the phase curve (Fig. 9) one may see the reflection effect. Such short period is quite uncharacteristic for Algol-type binaries, however the photometric features of the phase curve suggest that this star is EA-type.

\bigskip
\textit{4.5 VSX J053126.8+732909}

USNO-B1.0 1634-0053184 = VSX J053126.8+732909 is the only RRab-type variable star in our list. Its location is marked in Fig. 10. The period of this star is $P=0\pdot53094\pm0\pdot00012$, and the amplitude is $\Delta_m=0\mdot411\pm0\mdot014$ (min $=15\mdot697\pm0\mdot011$, max $=15\mdot286\pm0\mdot008$). The asymmetry of $A=0.322\pm0.015$ is clearly visible on the phase curve (Fig. 11.) These parameters are quite characteristic for RRab pulsating stars.

\bigskip
\textit{4.6 VSX J053300.0+732726}

The variable star USNO-B1.0 1634-0053325 = VSX J053300.0+732726 is again a binary system. Its position is marked in Fig. 12, the phase curve is shown in Fig. 13. The period of this star is $P=0\pdot34545\pm0\pdot00005$. The magnitudes in both minima were determined using MCV software by approximating the phase curve in minima with the algebraic polynomials of the optimal degree $s=6$: min$_{\rm I}=15\mdot853\pm0\mdot010$ and min$_{\rm II}=15\mdot795\pm0\mdot012$. The difference between minima $\Delta_m=0\mdot058\pm0\mdot016$ is small, so we may classify this binary as EW-type.

\bigskip
\textit{4.7 VSX J053444.4+734006}

The variable star USNO-B1.0 1636-0050887=VSX J053444.4+734006 was recognized as a close binary system of EW-type (W UMa) with the period $P=0\pdot29252\pm0\pdot00002$. The location of this variable is indicated in Fig. 14, and the phase curve is shown in Fig. 15. All photometric parameters were calculated using FDCN software. The depths of primary and secondary minima are nearly equal: min$_{\rm I}=13\mdot826\pm0\mdot002$, min$_{\rm II}=13\mdot817\pm0\mdot002$. From the shape of the phase curve, one may tentatively conclude that the components of this binary star significantly overfill its Roche lobes. However, for a more precise solution, spectral and multicolour observations are needed.

\bigskip
\textit{4.8 VSX J053514.5+733124}

The period of USNO-B1.0 1635-0051301 = VSX J053514.5+733124 is $P=0\pdot45692\pm0\pdot00003$. The position of this variable star is marked in Fig. 16. It is a binary system with nearly equal minima on the phase curve (Fig. 17): min$_{\rm I}=14\mdot339\pm0\mdot003$, min$_{\rm II}=14\mdot328\pm0\mdot005$. We classified this star as EW-type binary. The features of the phase curve suggest that the components of this binary star significantly overfill its Roche lobes. A slight O'Connell effect is noticable, max${\rm _I}=14\mdot105\pm0\mdot003$ and max$_{\rm II}=14\mdot094\pm0\mdot004$, thus the difference between brightness in maxima is ${\rm max_{\rm I}}-{\rm max_{{\rm II}}}=0\mdot011\pm0\mdot005$.

\bigskip
\textit{4.9 VSX J053638.2+731700}

The star USNO-B1.0 1632-0061720 = VSX J053638.2+731700 varies with the period $P=0\pdot35116\pm0\pdot00005$. The location of this star is shown in Fig. 18. The phase curve (Fig. 19) of this star is nearly symmetric. The magnitudes in maximum and minimum are max $=15\mdot504\pm0\mdot006$, min $=15\mdot873\pm0\mdot013$. Taking into account the period and the photometric features of the phase curve, we classified this variable as RRc-type star.

\bigskip
\textbf{5. Acknowledgements}

This research is based on data collected with the Tzec Maun Observatory, operated by the Tzec Maun Foundation. Special thanks to Ron Wodaski (director of the observatory) and Donna Brown-Wodaski (director of the Tzec Maun Foundation).

We would like to thank Tom Krajci for reviewing and correcting this paper. 

Also we are thankful to Ivan L. Andronov for helpful discussion.

This publication has been made using of the Aladin interactive sky atlas, operated at CDS, Strasbourg, France and the International Variable Star Index (VSX) maintained by AAVSO.

\bigskip
\textbf{References}
%\bigskip

%\begin{thebibliography} \\
\noindent Andronov, I.L. 1994, $OAP$, 7, 49 \\
Andronov, I.L. 2003, $ASPC$, 292, 391 \\
Andronov, I.L., and Baklanov, A.V. 2004, $Astronomy~ School~ Reports~$, 5, 264,\\
 \indent http://uavso.pochta.ru/mcv \\
Bertin, E., and Arnouts, S. 1996, $A\&AS$, 117, 393 \\
Lafler, J., and Kinman, T.D. 1965, $ApJ.Suppl.$, 11, 216 \\
Monet, D. G. et al. 2003, $AJ$, 125, 984 \\
Motl, D. 2007, C-Munipack Project v1.1,  http://integral.physics.muni.cz/cmunipack/index.html \\
Vanmunster, T. 2010, $PERANSO$, period analysis software,  http://www.peranso.com \\
Virnina, N. 2010, $OEJV$, 124, 1 \\
%\end{thebibliography}

%%%%%%%%%%%%%%%%%%%%%%%%%%%%%%%%

\begin{figure}[H]
\label{fig1}
\centering
\includegraphics[width=0.47\textwidth]{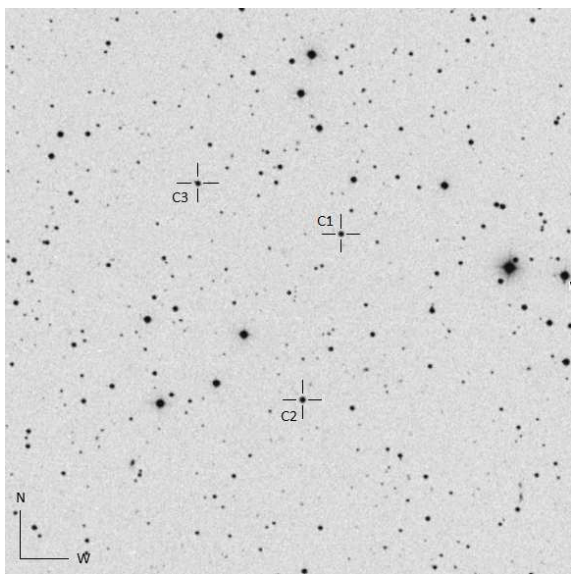}
\caption{\small Comparison stars, FOV is $15'\times15'$}
\end{figure}

\begin{figure}[H]

\hspace{0.05\textwidth}
\begin{minipage}[t]{0.32\textwidth}
\label{fig2}
\centering
\includegraphics[width=\textwidth]{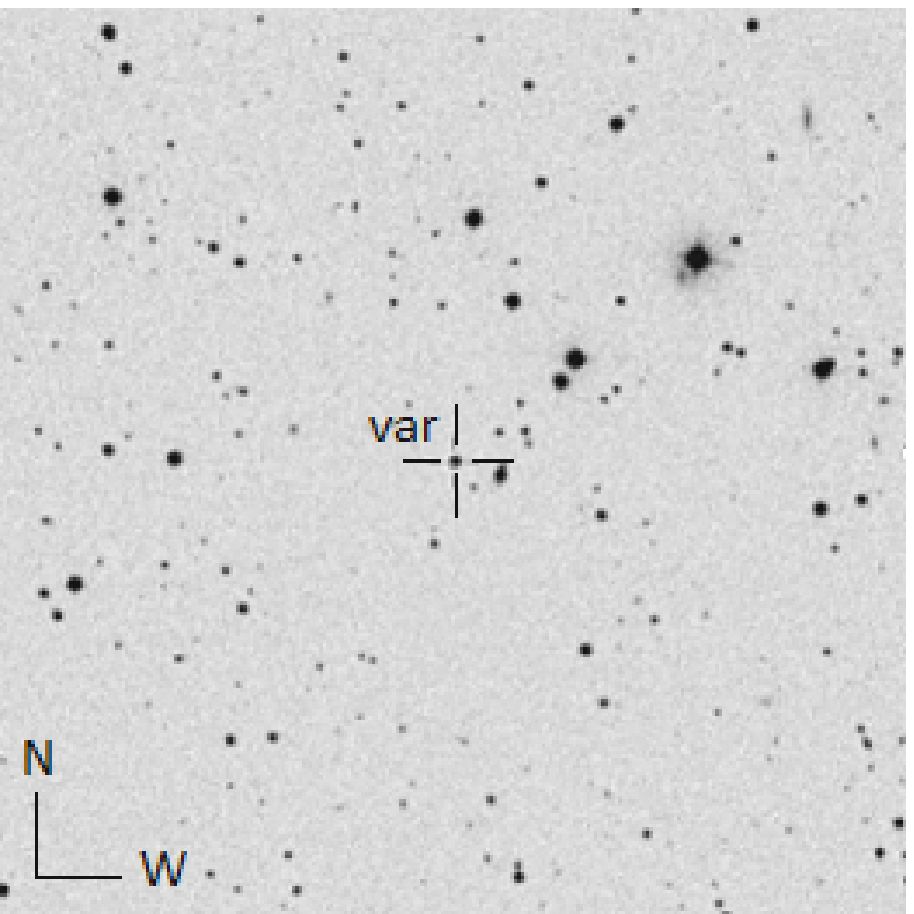}
\caption{\small Location of the variable VSX J052543.6+722840 ($10'\times10'$)}
\end{minipage}
\hspace{0.05\textwidth}
\begin{minipage}[t]{0.63\textwidth}
\label{fig3}
\centering
\includegraphics[width=\textwidth]{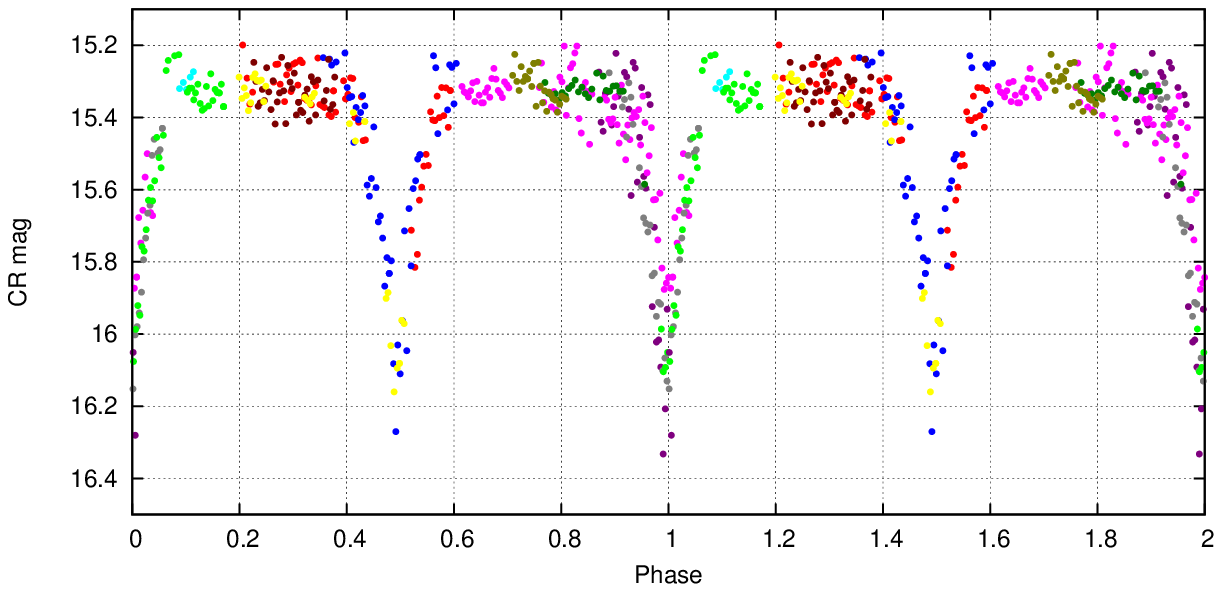}
\caption{\small Phase curve of EA-type variable VSX J052543.6+722840 ($P=0.64221(13)$d)}
\end{minipage}

\end{figure}

\begin{figure}[H]

\hspace{0.05\textwidth}
\begin{minipage}[t]{0.32\textwidth}
\label{fig4}
\centering
\includegraphics[width=\textwidth]{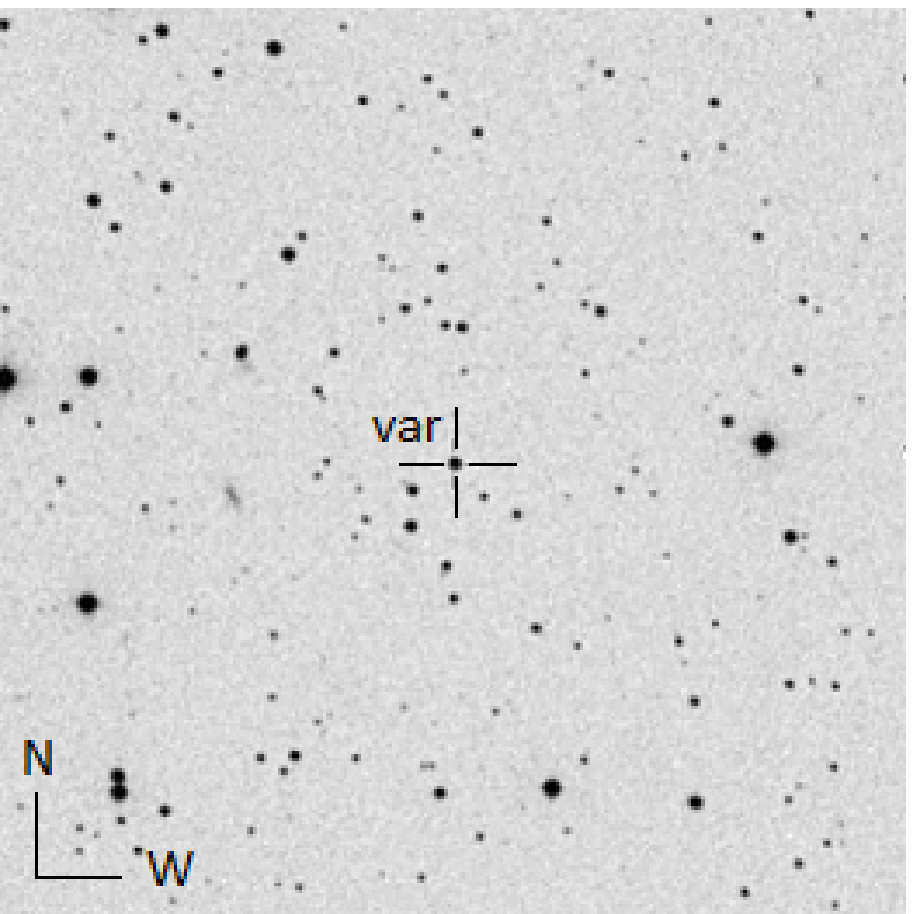}
\caption{\small Location of the variable VSX J052825.5+732114 ($10'\times10'$)}
\end{minipage}
\hspace{0.05\textwidth}
\begin{minipage}[t]{0.63\textwidth}
\label{fig5}
\centering
\includegraphics[width=\textwidth]{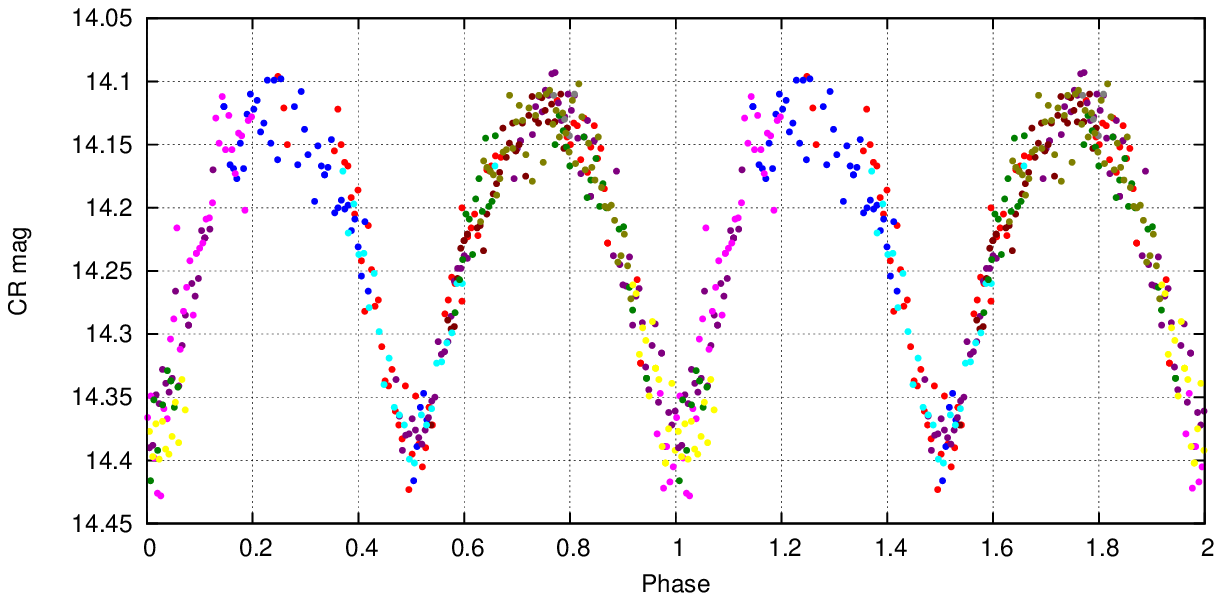}
\caption{\small Phase curve of EW-type variable VSX J052825.5+732114 ($P=0.41535(02)$d)}
\end{minipage}

\end{figure}

\begin{figure}[H]

\hspace{0.05\textwidth}
\begin{minipage}[t]{0.32\textwidth}
\label{fig6}
\centering
\includegraphics[width=\textwidth]{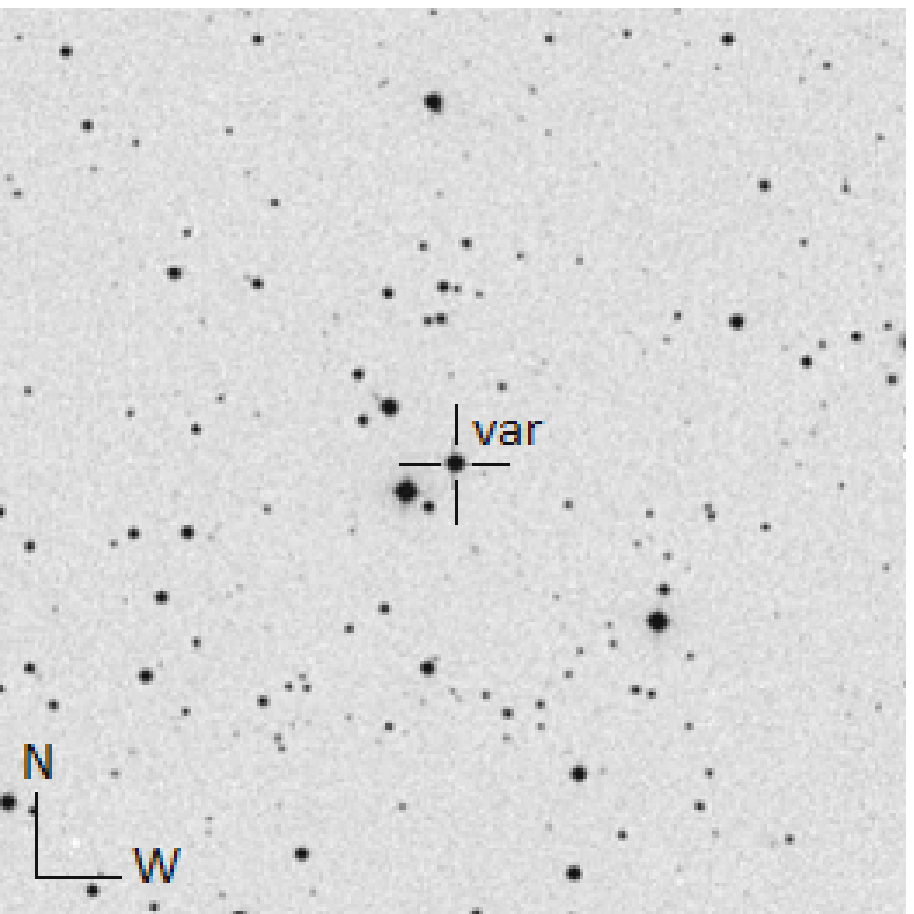}
\caption{\small Location of the variable VSX J052916.7+723300 ($10'\times10'$)}
\end{minipage}
\hspace{0.05\textwidth}
\begin{minipage}[t]{0.63\textwidth}
\label{fig7}
\centering
\includegraphics[width=\textwidth]{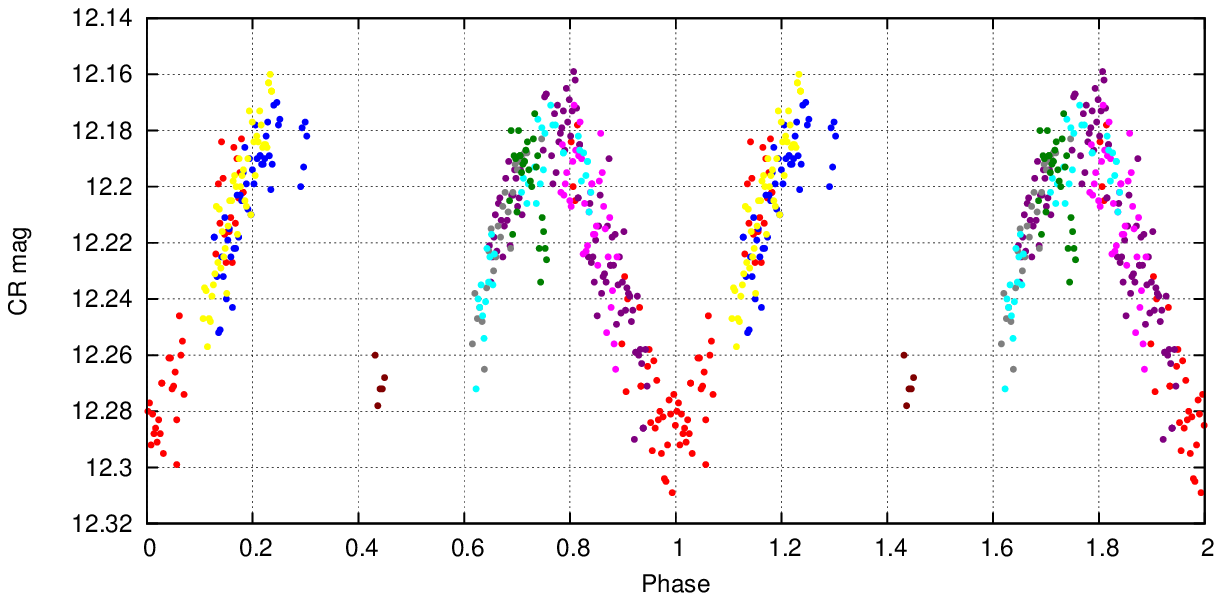}
\caption{\small Phase curve of Ell-type variable VSX J052916.7+723300 ($P=0.91252(15)$d)}
\end{minipage}

\end{figure}

\begin{figure}[H]

\hspace{0.05\textwidth}
\begin{minipage}[t]{0.32\textwidth}
\label{fig8}
\centering
\includegraphics[width=\textwidth]{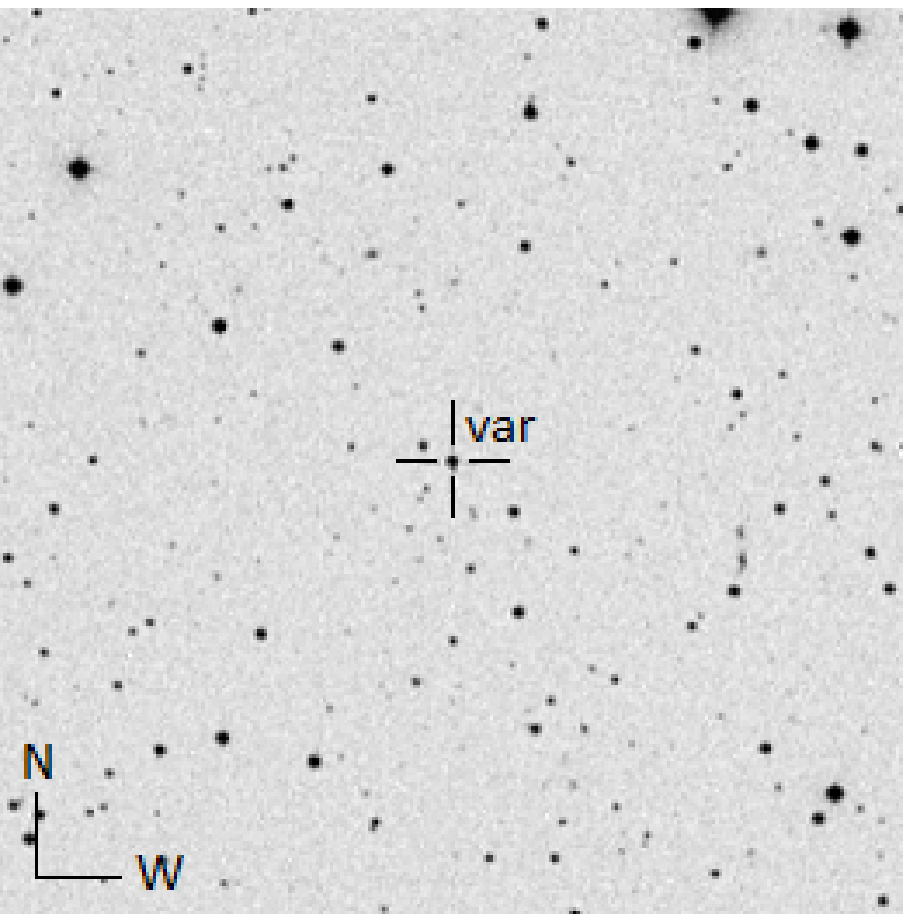}
\caption{\small Location of the variable VSX J053044.4+725113 ($10'\times10'$)}
\end{minipage}
\hspace{0.05\textwidth}
\begin{minipage}[t]{0.63\textwidth}
\label{fig9}
\centering
\includegraphics[width=\textwidth]{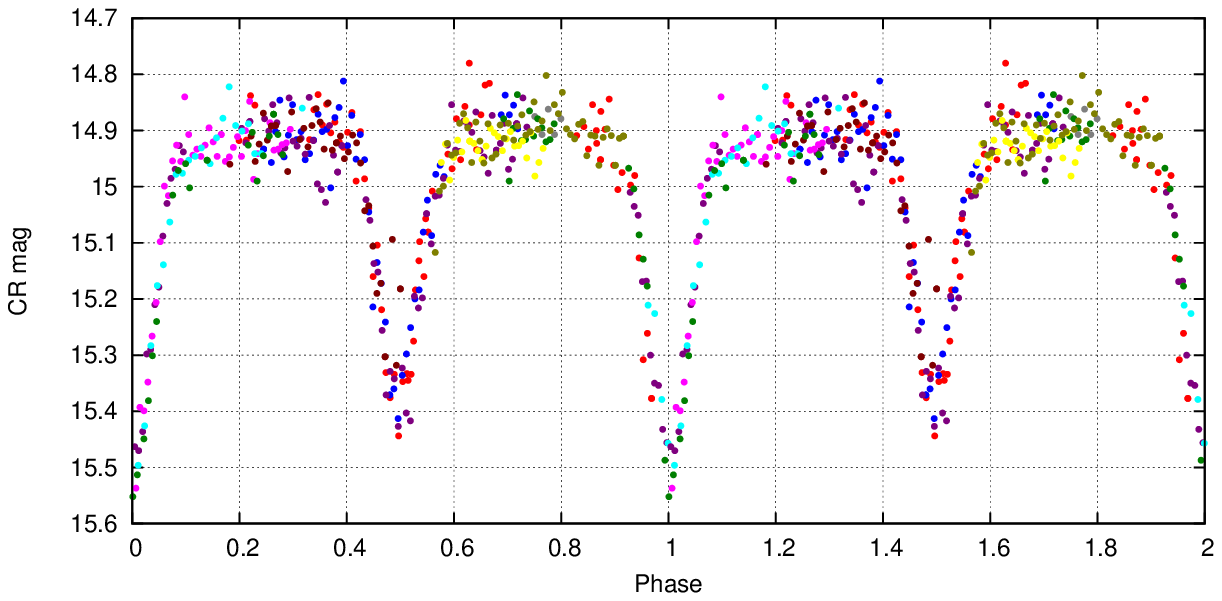}
\caption{\small Phase curve of EA-type variable VSX J053044.4+725113 ($P=0.33648(02)$d)}
\end{minipage}

\end{figure}

\begin{figure}[H]

\hspace{0.05\textwidth}
\begin{minipage}[t]{0.32\textwidth}
\label{fig10}
\centering
\includegraphics[width=\textwidth]{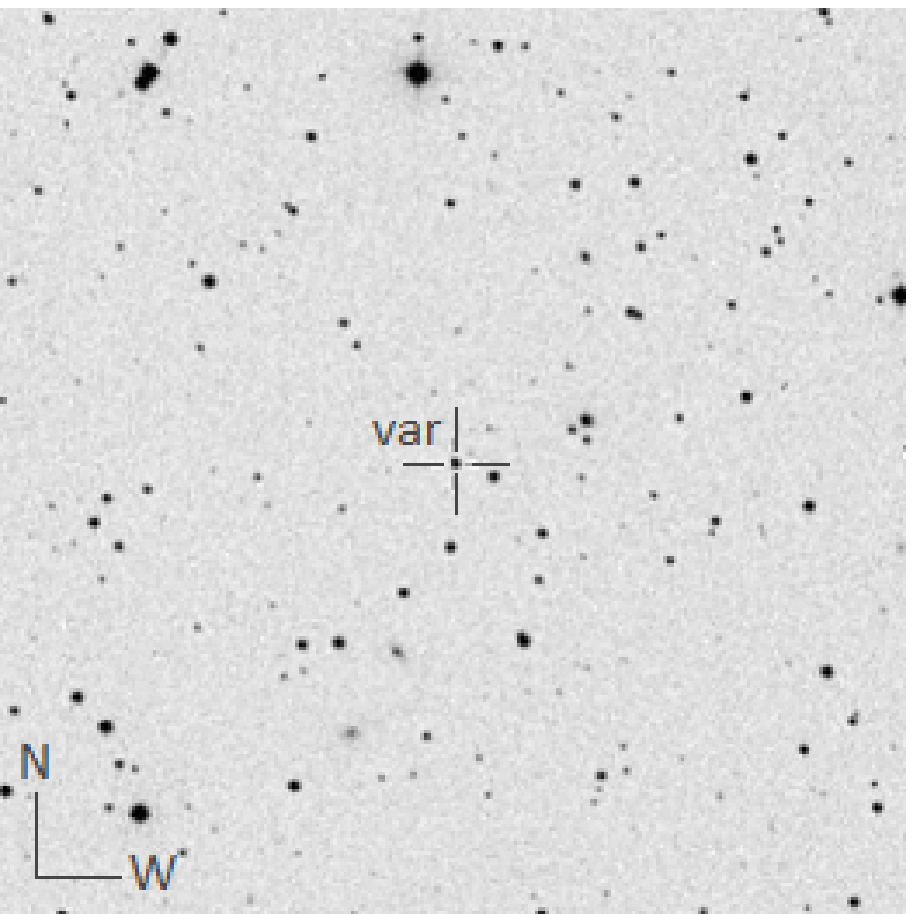}
\caption{\small Location of the variable VSX J053126.8+732909 ($10'\times10'$)}
\end{minipage}
\hspace{0.05\textwidth}
\begin{minipage}[t]{0.63\textwidth}
\label{fig11}
\centering
\includegraphics[width=\textwidth]{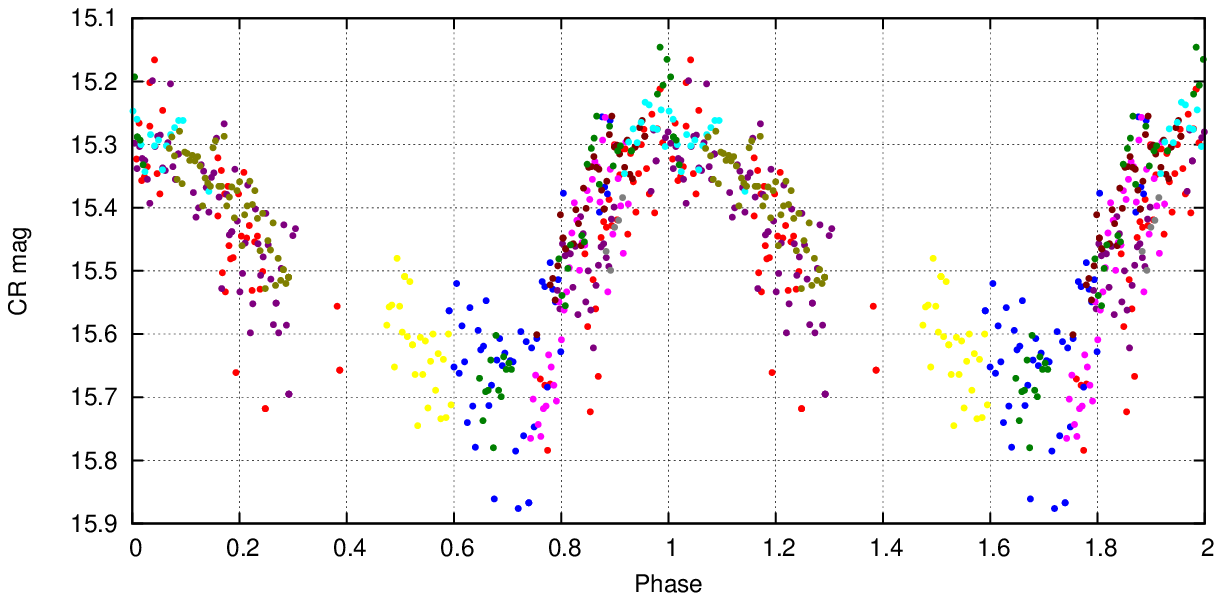}
\caption{\small Phase curve of RRab-type variable VSX J053126.8+732909 ($P=0.53094(12)$d)}
\end{minipage}

\end{figure}

\begin{figure}[H]

\hspace{0.05\textwidth}
\begin{minipage}[t]{0.32\textwidth}
\label{fig12}
\centering
\includegraphics[width=\textwidth]{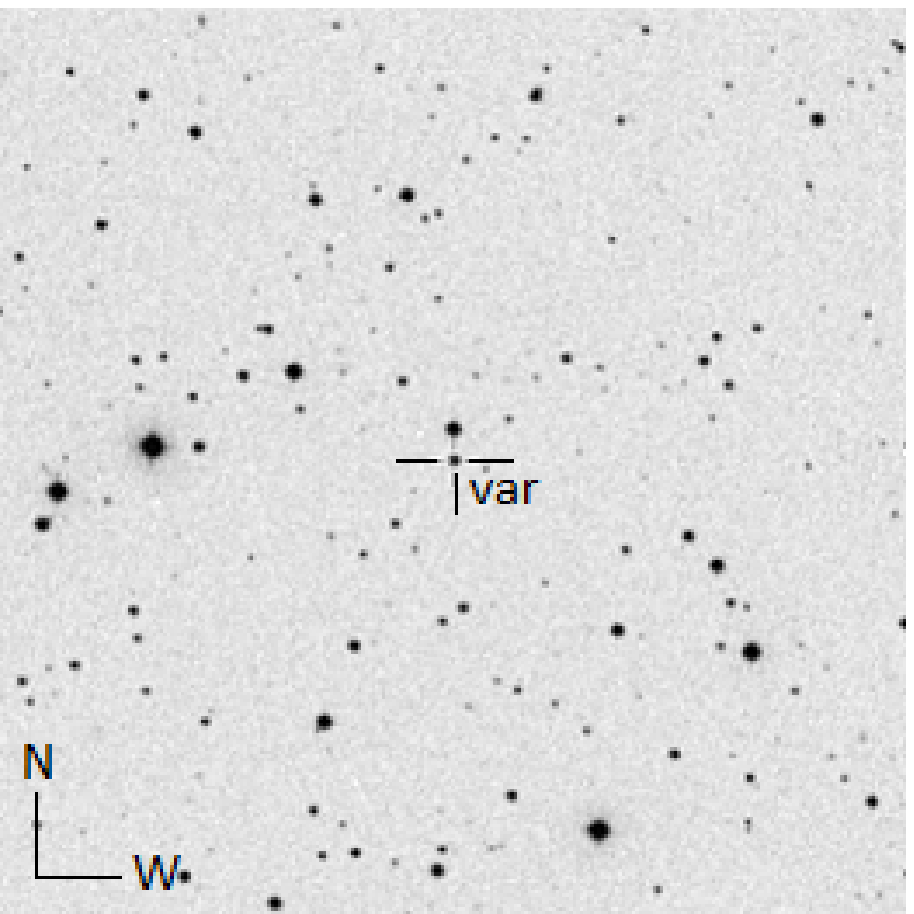}
\caption{\small Location of the variable VSX J053300.1+732726 ($10'\times10'$)}
\end{minipage}
\hspace{0.05\textwidth}
\begin{minipage}[t]{0.63\textwidth}
\label{fig13}
\centering
\includegraphics[width=\textwidth]{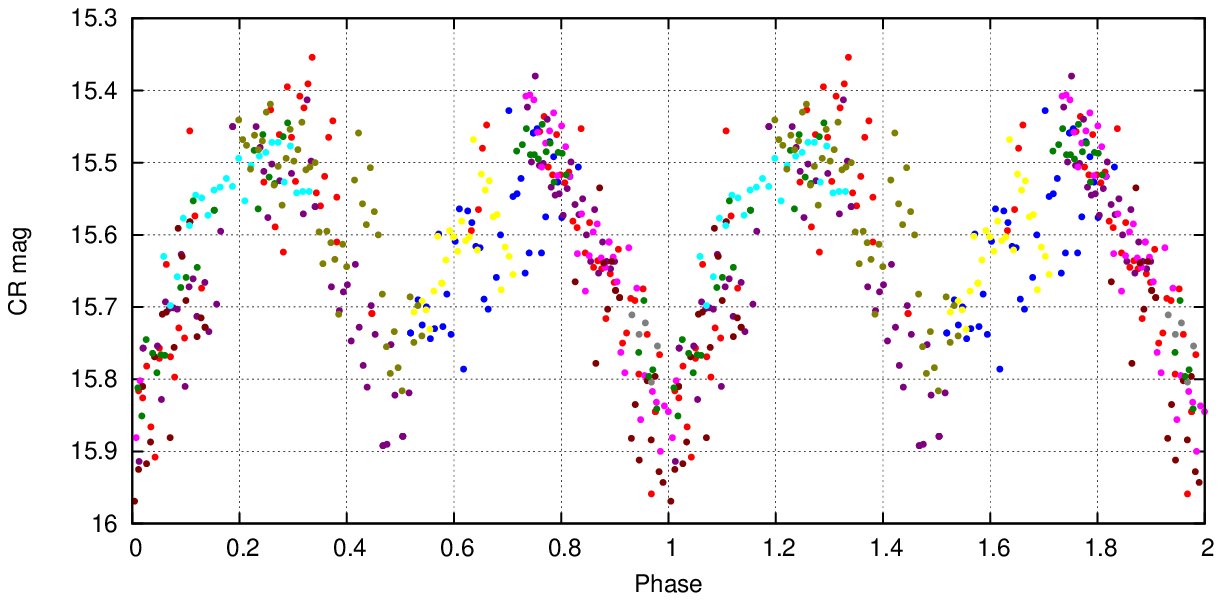}
\caption{\small Phase curve of EW-type variable VSX J053300.1+732726 ($P=0.34545(05)$d)}
\end{minipage}

\end{figure}

\begin{figure}[H]

\hspace{0.05\textwidth}
\begin{minipage}[t]{0.32\textwidth}
\label{fig14}
\centering
\includegraphics[width=\textwidth]{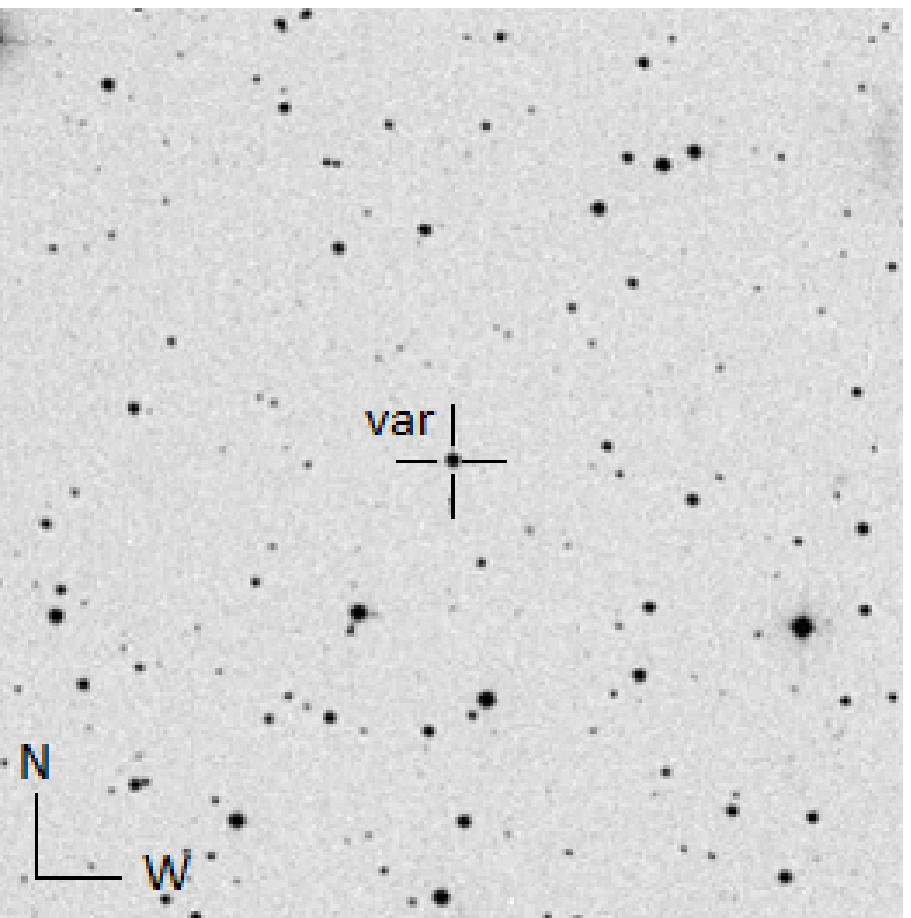}
\caption{\small Location of the variable VSX J053444.4+734006 ($10'\times10'$)}
\end{minipage}
\hspace{0.05\textwidth}
\begin{minipage}[t]{0.63\textwidth}
\label{fig15}
\centering
\includegraphics[width=\textwidth]{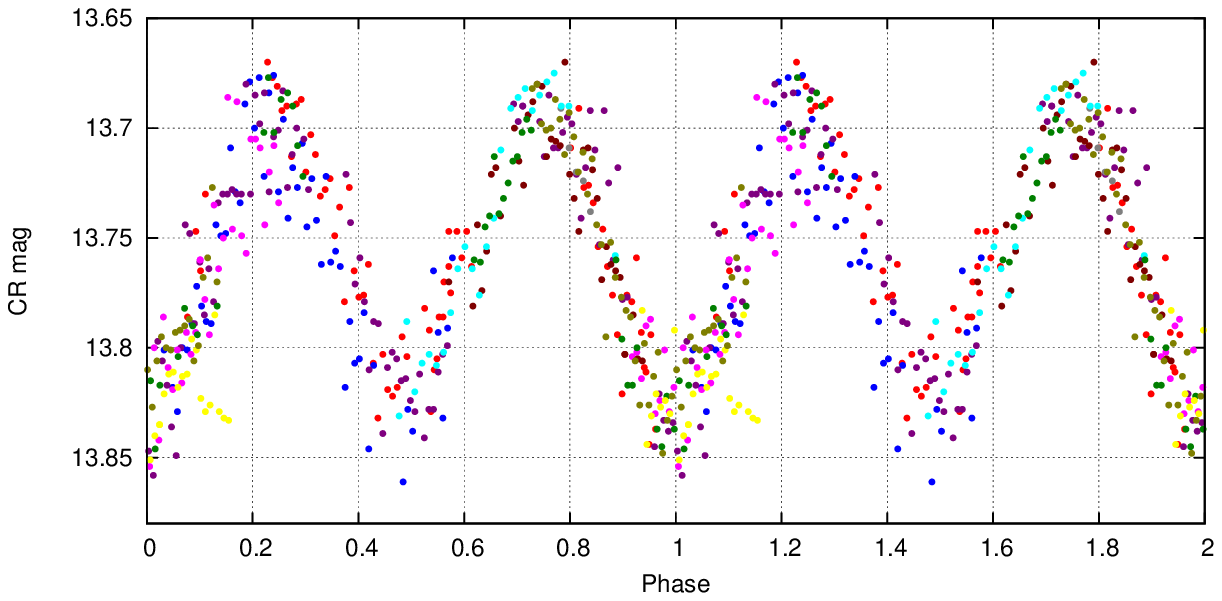}
\caption{\small Phase curve of EW-type variable VSX J053444.4+734006 ($P=0.29252(02)$d)}
\end{minipage}

\end{figure}

\begin{figure}[H]

\hspace{0.05\textwidth}
\begin{minipage}[t]{0.32\textwidth}
\label{fig16}
\centering
\includegraphics[width=\textwidth]{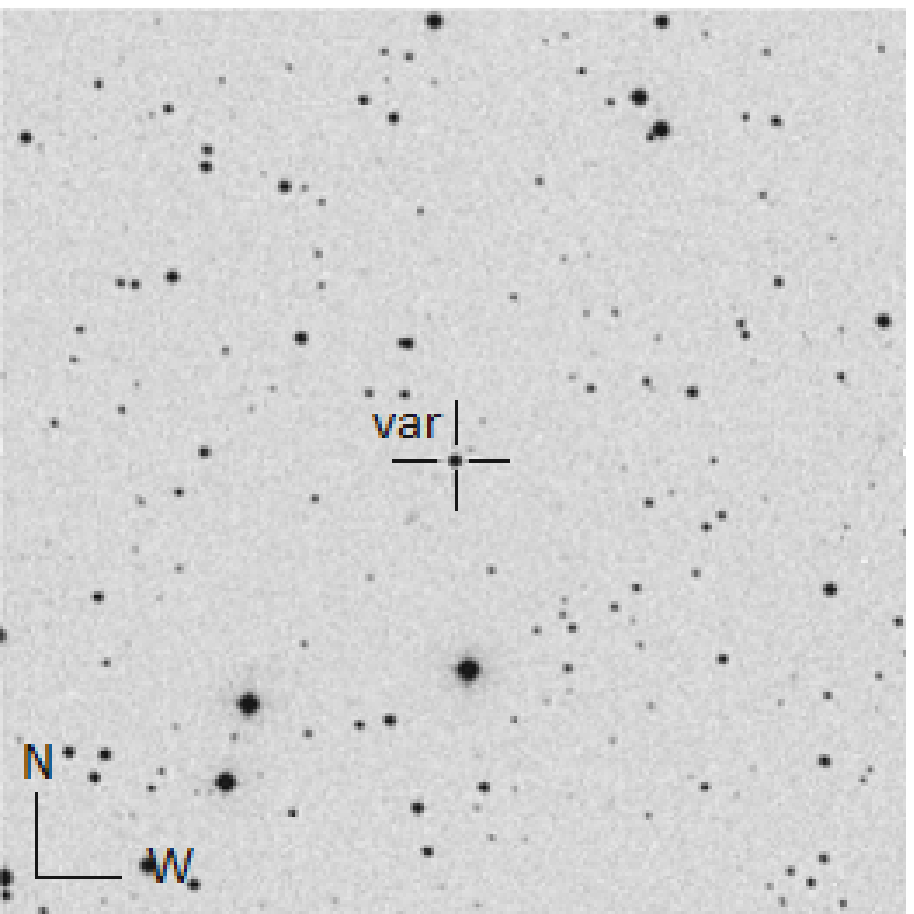}
\caption{\small Location of the variable VSX J053514.5+733124 ($10'\times10'$)}
\end{minipage}
\hspace{0.05\textwidth}
\begin{minipage}[t]{0.63\textwidth}
\label{fig17}
\centering
\includegraphics[width=\textwidth]{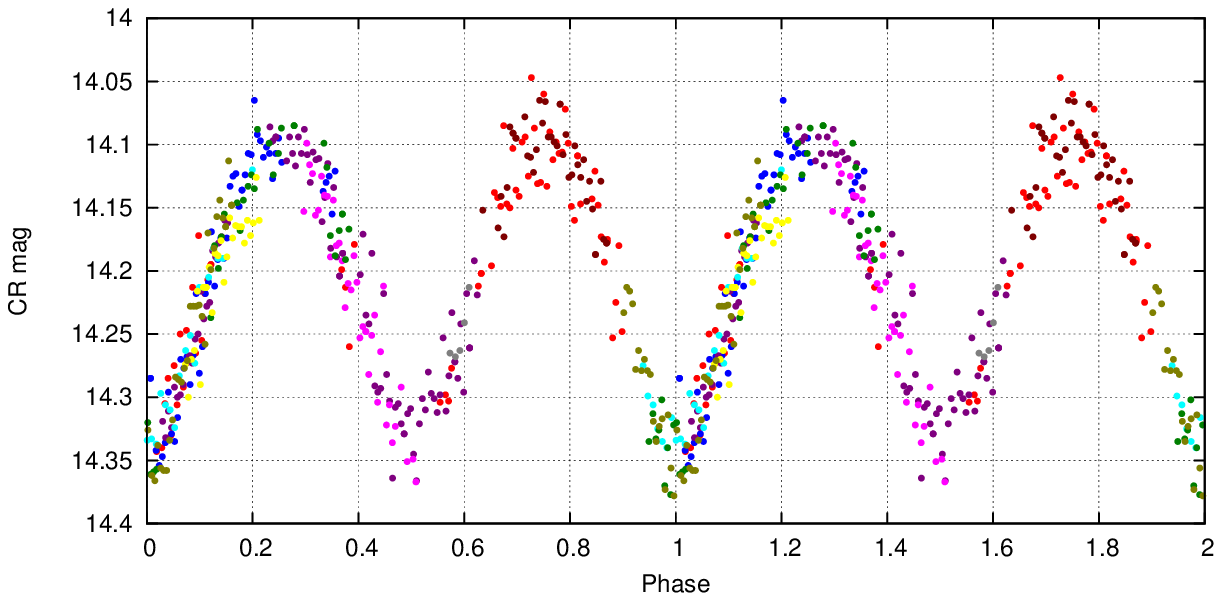}
\caption{\small Phase curve of EW-type variable VSX J053514.5+733124 ($P=0.45692(03)$d)}
\end{minipage}

\end{figure}

\begin{figure}[H]

\hspace{0.05\textwidth}
\begin{minipage}[t]{0.32\textwidth}
\label{fig18}
\centering
\includegraphics[width=\textwidth]{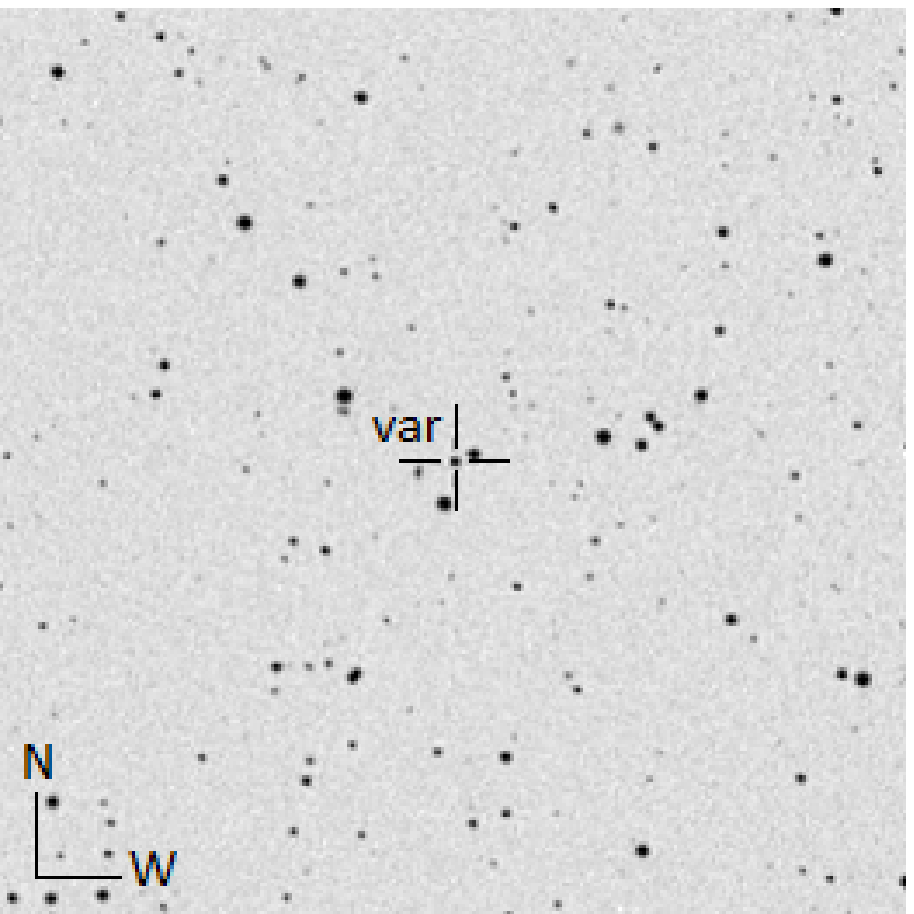}
\caption{\small Location of the variable VSX J053638.3+731700 ($10'\times10'$)}
\end{minipage}
\hspace{0.05\textwidth}
\begin{minipage}[t]{0.63\textwidth}
\label{fig19}
\centering
\includegraphics[width=\textwidth]{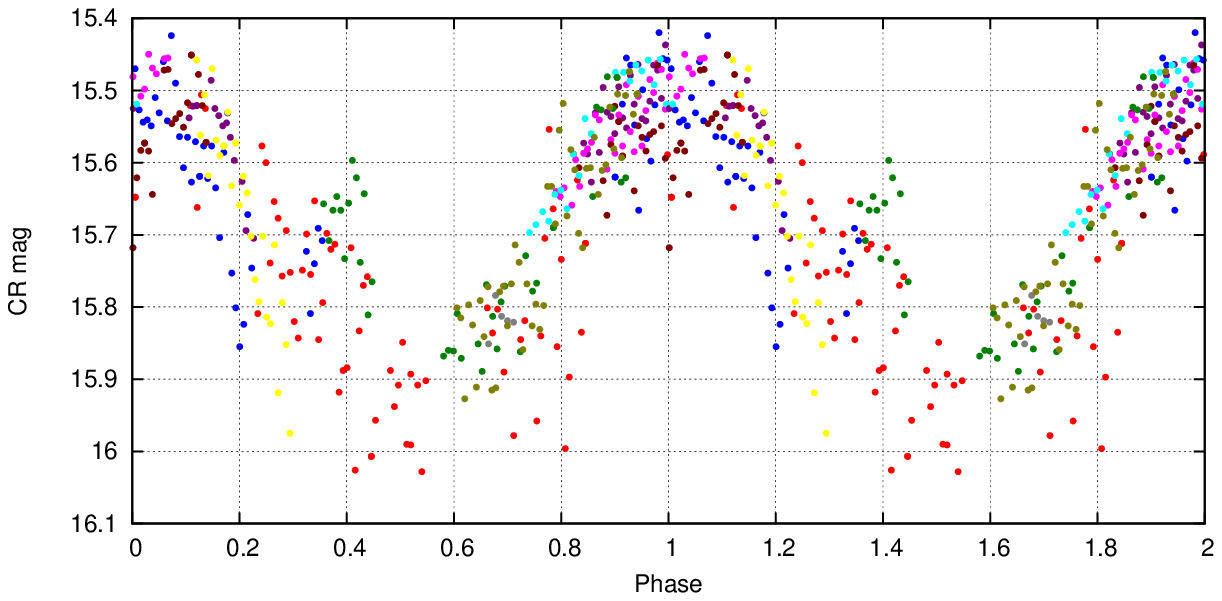}
\caption{\small Phase curve of RRc-type variable VSX J053638.3+731700 ($P=0.35116(05)$d)}
\end{minipage}

\end{figure}

%%%%%%%%%%%%%%%%%%%%%%%%%%%%%%%%%%%%%%%%%%%%%%%%%%%%%

\begin{table}[H]
\centering
\label{table1}
\caption{ Magnitudes of reference stars in the standard system measured by APASS (AAVSO)}
\begin{tabular}{ccccccc}
\hline
No & USNO-B1.0   &    $\alpha(2000)$    &     $\delta(2000)$  & $B, m$ &  $V, m$   &   $R_c, m$\\ 
\hline
1 & 1623-0066292 & $05^h 35^m 05.658^s$ & $+72^\circ 18' 02.91''$ & 15.145 & 14.343 & 13.832 \\
2 & 1622-0068625 & $05^h 33^m 29.060^s$ & $+72^\circ 17' 02.95''$ & 15.411 & 14.496 & 13.987 \\
3 & 1622-0068584 & $05^h 33^m 03.919^s$ & $+72^\circ 17' 32.81''$ & 15.507 & 14.865 & 14.400 \\
4 & 1622-0068513 & $05^h 32^m 29.179^s$ & $+72^\circ 17' 25.85''$ & 14.001 & 13.320 & 12.861 \\
5 & 1622-0067988 & $05^h 29^m 09.585^s$ & $+72^\circ 17' 14.73''$ & 14.607 & 13.560 & 12.991 \\
\hline
\end{tabular}
\end{table}

\begin{table}[H]
\centering
\label{table2}
\caption{ Instrumental $R$ magnitudes of comparison stars}
\begin{tabular}{ccccc}
\hline
No & USNO-B1.0   &    $\alpha(2000)$    &     $\delta(2000)$  & $R$, m\\
\hline
1 & 1629-0065214 & $05^h 31^m 05.213^s$ & $+72^\circ 57' 01.60''$ & 13.985(27)\\
2 & 1628-0064903 & $05^h 31^m 18.911^s$ & $+72^\circ 52' 40.71''$ & 13.466(19)\\
3 & 1629-0065309 & $05^h 31^m 56.236^s$ & $+72^\circ 58' 20.81''$ & 14.044(22)\\
\hline
\end{tabular}
\end{table}

\begin{table}[H]
\centering
\label{table3}
\caption{ Cross-identification of new variable stars}
\begin{tabular}{ccccc}
\hline
No & USNO-B1.0   &    $\alpha(2000)$    &     $\delta(2000)$  & VSX\\
\hline
1 & 1624-0065083 & $05^h 25^m 43.672^s$ & $+72^\circ 28' 40.10''$ & J052543.6+722840	\\
2 & 1633-0056300 & $05^h 28^m 25.539^s$ & $+73^\circ 21' 14.57''$ & J052825.5+732114	\\
3 & 1625-0065763 & $05^h 29^m 16.741^s$ & $+72^\circ 33' 00.63''$ & J052916.7+723300	\\
4 & 1628-0064829 & $05^h 30^m 44.331^s$ & $+72^\circ 51' 13.67''$ & J053044.4+725113	\\
5 & 1634-0053184 & $05^h 31^m 26.847^s$ & $+73^\circ 29' 09.88''$ & J053126.8+732909	\\
6 & 1634-0053325 & $05^h 33^m 00.072^s$ & $+73^\circ 27' 26.52''$ & J053300.1+732726	\\
7 & 1636-0050887 & $05^h 34^m 44.439^s$ & $+73^\circ 40' 06.37''$ & J053444.4+734006	\\
8 & 1635-0051301 & $05^h 35^m 14.565^s$ & $+73^\circ 31' 24.19''$ & J053514.5+733124	\\
9 & 1632-0061720 & $05^h 36^m 38.260^s$ & $+73^\circ 17' 00.29''$ & J053638.3+731700	\\
\hline
\end{tabular}
\end{table}

\begin{table}[H]
\centering
\label{table4}
\caption{Photometric parameters of new variable stars}
\begin{tabular}{clccccc}
\hline
No&	type&  $max$&     min$_{\rm I}$&   min$_{\rm II}$&       period &    $HJD_0,~ 2455$\\
\hline
1& EA  & 15.318(14)& 16.035(26) &    15.987(21) &    0.64221(13)&  261.1514(14)\\
2& EW  & 14.125(03)& 14.386(03) &    14.383(05) &    0.41535(02)&  261.9159(06)\\
3& Ell & 12.177(02)& \multicolumn{2}{c}{12.292(04)}& 0.91252(15)&  251.7278(23)\\
4& EA  & 14.901(06)& 15.476(10) &    15.369(10) &    0.33648(02)&  261.9981(05)\\
5& RRab& 15.286(08)& \multicolumn{2}{c}{15.697(11)}& 0.53094(12)&  261.8534(55)\\
6& EW  & 15.468(10)& 15.853(10) &    15.795(12) &    0.34545(05)&  262.1286(12)\\
7& EW  & 13.691(02)& 13.826(02) &    13.817(02) &    0.29252(02)&  261.8484(05)\\
8& EW  & 14.094(04)& 14.339(03) &    14.328(05) &    0.45692(03)&  261.8896(10)\\
9& RRc & 15.504(06)& \multicolumn{2}{c}{15.873(13)}& 0.35116(05)&  253.7041(95)\\
\hline
\end{tabular}
\end{table}

\end{document}